\documentclass[fleqn,usenatbib]{mnras}

\usepackage{newtxtext,newtxmath}

\usepackage{xcolor}
\usepackage{cancel}
\usepackage{ae,aecompl}
\usepackage[T1]{fontenc}
\usepackage{graphicx}	
\usepackage{amsmath}	

\usepackage{amsmath}	

\usepackage{amsmath}	
\DeclareRobustCommand{\VAN}[3]{#2}
\let\VANthebibliography\thebibliography
\def\thebibliography{\DeclareRobustCommand{\VAN}[3]{##3}\VANthebibliography}

\title[Brightest Stream Progenitors]{Linking the brightest stellar streams with the accretion history of Milky Way-like galaxies}

\author[A. Vera-Casanova et al.]{
Alex Vera-Casanova$^{1}$\thanks{E-mail: alex.vera@userena.cl},
Facundo A. G{\'o}mez$^{1,2}$,
Antonela Monachesi$^{1,2}$,\newauthor
Ignacio Gargiulo$^{3,4}$,
Diego Pallero$^{1}$,
Robert J. J. Grand$^{5}$,
Federico Marinacci$^{6}$,\newauthor
R\"udiger Pakmor$^{5}$,
Christine M. Simpson$^{7,8}$,
Carlos S. Frenk$^{9}$,
and Gustavo Morales$^{1}$
\\
$^{1}$Departamento de Astronom\'ia, Universidad de La Serena, Avenida Juan Cisternas 1200, La Serena, Chile\\
$^{2}$Instituto de Investigaci\'on Multidisciplinar en Ciencia y Tecnolog\'ia, Universidad de La Serena, Ra\'ul Bitr\'an 1305, La Serena, Chile\\
$^{3}$ Instituto de Astrof{\'i}sica de La Plata (CCT La Plata, CONICET, UNLP), Paseo del Bosque s/n, B1900FWA, La Plata, Argentina\\
$^{4}$ Consejo Nacional de Investigaciones Cient{\'i}ficas y T{\'e}cnicas, Rivadavia 1917, C1033AAJ Buenos Aires, Argentina\\
$^{5}$Max-Planck-Institut f\"ur Astrophysik, Karl-Schwarzschild-Str. 1, 85748 Garching, Germany\\
$^{6}${Department of Physics \& Astronomy "Augusto Righi", University of Bologna, via Gobetti 93/2, 40129 Bologna, Italy}\\
$^{7}$Department of Astronomy \& Astrophysics, The University of Chicago, Chicago, IL 60637, USA\\
$^{8}$Enrico Fermi Institute, The University of Chicago, Chicago, IL 60637, USA\\
$^{9}$Institute for Computational Cosmology, Department of Physics, University of Durham, South Road, Durham DH1 3LE, UK\\
}

\date{Accepted XXX. Received YYY; in original form ZZZ}

\pubyear{2022}

\begin{document}
\label{firstpage}
\pagerange{\pageref{firstpage}--\pageref{lastpage}}
\maketitle

\begin{abstract}
According to the current galaxy formation paradigm, mergers and interactions play an important role in shaping present-day galaxies. The remnants of this merger activity can be used to constrain galaxy formation models. In this work we use a sample of thirty hydrodynamical simulations of Milky Way-mass halos, from the AURIGA project, to generate surface brightness maps and search for the brightest stream in each halo as a function of varying limiting magnitude. We find that none of the models shows signatures of stellar streams at $\mu_{r}^{lim} \leq 25$ mag arcsec$^{-2}$. The stream detection increases significantly between 28 and 29 mag arcsec$^{-2}$. Nevertheless, even at 31 mag arcsec$^{-2}$, 13  percent of our models show no detectable streams. We study the properties of the brightest streams progenitors (BSPs). We find that BSPs are accreted within a broad range of infall times, from 1.6 to 10 Gyr ago, with only 25 percent accreted within the last 5 Gyrs; thus most BSPs correspond to relatively early accretion events. We also find that 37 percent of the BSPs survive to the present day. The median infall times for surviving and disrupted BSPs are 5.6 and 6.7 Gyr, respectively. We find a clear relation between  infall time and infall mass of the BSPs, such that more massive progenitors tend to be accreted at later times. However, we find that the BSPs are not, in most cases, the dominant contributor to the accreted stellar halo of each galaxy.

\end{abstract}




\section{Introduction}

A well-tested prediction from the current paradigm of galaxy formation establishes that galaxies grow in mass by the accretion of material from the surrounding environment \citep[e.g.][]{1991ApJ...379...52W}. In addition to mass growth, the accretion of satellites plays a fundamental role in shaping the properties of the galaxies we observed at the present-day. The interaction and merger with massive objects can induce a wide variety of perturbations in the central galactic regions. These can range from the destruction of pre-existing discs in the most extreme case to the excitation of non-axisymmetric perturbations such as bars, spirals, warps and lopsidedness \citep[e.g.,][]{2009PhR...471...75J,2009MNRAS.397.1599Q,2016MNRAS.456.2779G,2016MNRAS.459..199G,2017MNRAS.465.3446G,2020arXiv201112323G}. Smaller satellites, i.e. those with host-to-satellite mass ratios $\lesssim$ 1:10, are less likely to imprint lasting and global perturbations in the inner galactic regions. However, they significantly contribute to the formation of the outer spheroidal and extended galactic component of galaxies, known as the stellar halo \citep[e.g.,][]{Searle_Zinn78, White_Rees78, 2005_Bullock_Johnston}. Low-mass satellites are not as strongly affected by dynamical friction as their more massive counterparts. As a result, these objects can spend long periods in the outer galactic region as they are tidally disrupted, leaving behind extended low surface brightness substructures known as tidal streams \citep[e.g.,][]{Johnston96, 1999Majewski, 2009Natur.461...66M, 2010Marinez-Delgado}. 

Substructure left in halos by satellites of any mass are considered fossil signatures of accretion events since they can provide detailed information about the merging history of the host galaxy. As such, streams are being actively searched for not only in the Milky Way, but also in nearby galaxy using different techniques. In the Milky Way it is possible to use measurements of the full six-dimensional phase-space of stars. This make it possible even to identify substructures in the inner galactic region, where the mixing times are short and streams are typically well mixed in configuration space. After the pioneering work of \citet{1999Helmi}, several studies have been dedicated at quantifying amount of substructures in the solar vicinity. Recently, thanks to the data from the astrometric satellite Gaia \citep{Gaia2018}, the number of known streams has widely grown \citep[e.g.][]{2020arXiv201205245I}. The combination of this information with the mapping of the outer halo with photometric and spectroscopic surveys is allowing us, for the first time, to obtain a comprehensive view of the merging history of our Galaxy \citep[e.g.,][]{H3,2019ApJ...887..237C, 2020ARA&A..58..205H,2020ApJ...901...48N}.

Information about individual stars located within a significant area of the stellar halo can only be obtained for nearby galaxies \citep{2014A&A...562A..73G, 2016ApJ...823...19Crno, 2020ApJ...905...60S}. Thus, for the vast majority of Milky Way-type galaxies, we rely on surface brightness maps obtained from integrated photometry. A number of observational surveys have capitalized on this technique to study the merging histories of several galaxies \citep[e.g.,][]{2010Marinez-Delgado,2013ApJ...765...28A,2014VanDokkum, 2018ApJ...866..103K, 2018A&A...614A.143M}. Thanks to  very deep observations, these studies have started to conduct a census of the  stellar streams in the nearby Universe. A common goal is to assess the frequency with which such extended stellar streams can be observed as a function of limiting surface brightness and, thus, to constrain the merging activity these galaxies have undergone. However, in spite of the long exposure times, integrated light observations of stellar halos have typically reached surface brightness levels of $\mu \lesssim 28~$ mag arcsec$^{-2}$ (although see \citealt{2016ApJ...830...62M, 2016ApJ...823..123T} for deeper observations of a few individual galaxies), and thus are typically able to detect only the brightest stellar streams.  

On the theoretical side, several work have used cosmological simulations to interpret both the integrated light observations of stellar halos \citep[][]{2018ApJ...869...12S, 2020MNRAS.495.4570M} as well as their individual stars \citep{2018NatAs...2..737D,2019MNRAS.485.2589M, 2021MNRAS.505..783F}. The analysis of simulations of stellar halos, in concert with observations, allows us to connect the observable quantities with the accretion history of galaxies \citep{2018NatAs...2..737D,2019MNRAS.485.2589M, 2017MNRAS.469L..48A} as well as to understand systematic in stellar halo properties obtained from observational methods \citep[e.g.][]{2018ApJ...869...12S} or to constraint the stellar mass-halo mass relation for dwarf galaxies \citep[e.g.][]{2022MNRAS.510.4208R}. In particular, we now know that stellar halos are primarily  built up from a few massive accretion events \citep[][]{2016ApJ...821....5D, 2017ApJ...837L...8B, 2019MNRAS.485.2589M, 10.1093/mnras/staa2221}, and that the information we extract from observations within 50 kpc mainly informs us about the properties of the most massive satellite accreted.

In this work, we use a suite of state-of-the-art cosmological hydrodynamical simulations of late-type galaxies from the Auriga project \citep{AURIGA} to analyze the information that can be extracted from the brightest stellar stream in each halo in regards to  their hosts merging activity. Our study builds up upon previous work by \citet[][hereafter J08]{2008ApJ...689..936J} who used cosmologically-motivated simulations to study how the frequency and properties of stellar halo substructure, as a function of surface brightness, are indicators of the recent merging histories of galaxies. Here we further study this problem by focusing on the brightest stellar streams of our simulated halos. This is of particular interest since it allows us to make a direct link between a specific and simple observable, i.e. the brightest stream of a galaxy halo, and the accretion history of the galaxy. Unlike previous studies who have used dark matter only simulations together with a particle tagging technique \citep[e.g. J08][]{cooper2010},  here we analyze high-resolution fully cosmological magneto-hydrodynamical simulations of the formation and evolution of late-type galaxies that naturally account for the different distributions of the satellite dark matter and stellar components.

Our paper is organized as follows. In Section \ref{sec:auriga} we introduce the Auriga simulations and discuss their suitability for this project. In Section \ref{sec:brightest stellar}, we discuss the generation of the surface brightness maps that are used to identify the brightest low surface brightness feature in each halo. We also quantify the number of models with observable streams as a function of limiting surface brightness.  In Section \ref{sec:properties}, we study the progenitor satellites from which each brightest stream originated, and characterize their distribution of infall times and infall mass by each progenitor. We present our Summary and Conclusions in Section \ref{sec:conclusion}.

\section{Auriga Simulations}
\label{sec:auriga}

The Auriga Project consists of a set of fifty cosmological magneto-hydrodynamic simulations of galaxies like the Milky Way \citep{AURIGA}. These are zoom-in simulations of dark matter halos chosen from the EAGLE project \citep{EAGLE}. The halos analyzed in this work were selected to have a narrow mass range of 1$ < M_{200}/10^{12}M_{\odot} < 2$, leaving us with a subset of thirty models. Each simulation was run assuming the $\Lambda$CDM cosmology, with parameters 
$\Omega_m$ = 0.307, $\Omega_b$ = 0.048, $\Omega_{\Lambda}$ = 0.693, and Hubble's constant $H_0$ = 100 \textit{h} km s$^{-1}$ Mpc$^{-1}$, \textit{h} $= 0.6777$ \citep{param_cosmo_planck}.
The multi-mass `zoom-in' resimulations were performed in a a periodic cube of side 100 \textit{h}$^{-1}$ Mpc using the N-body magneto-hydrodynamic moving mesh  code {\sc arepo} (\citealt{springel_2010}; \citealt{pakmor_2016}). 
For the simulations analyzed here, the dark matter and baryonic mass  is $\sim 3 \times 10^5 M_{\odot}$ and $\sim 5 \times 10^4 M_{\odot}$, respectively. The gravitational softening length for stellar and dark matter particles grows with scale factor up to a maximum of 369 pc. For the gas cells, the softening length scales with the mean radius of the cell but is not allowed to drop below the stellar softening length.

{\sc arepo} includes a  comprehensive galaxy formation model \citep{Vogelsberger2013}, including baryonic processes such as primordial and metal line cooling, a prescription for a uniform background UV field for reionization, a subgrid model for star formation \citep{Springel_Hernquist_2003}, a subgrid model for two-phase interstellar medium in pressure equilibrium \citep{Springel_Hernquist_2003},
magnetic fields \citep{pakmor_springel2013, pakmor2014}, gas accretion onto
black holes and energetic feedback from AGN and supernovae type
II (SNII) \citep[for more details see][]{Springel2005,  Vogelsberger2013, Marinacci2014, AURIGA}. The parameters that regulate the efficiency of each physical process were chosen by comparing the results obtained in simulations of cosmologically representative regions to a wide range of observations of the galaxy population.

In our models, each stellar particle represents a single
stellar population of a given mass, age and metallicity. Mass loss
and metal enrichment from type Ia supernovae (SNIa) and asymptotic giant branch (AGB) stars are modeled by calculating at each
time step the mass moving off the main sequence for each star particle according to a delay time distribution. Using the stellar population synthesis models from \citet{BruzualCharlot}, the luminosity of each stellar particle was estimated in multiple photometric bands. As a result, our models include detailed photometric luminosity estimates in the  U, B, V, g, r, i, z, K bands, all without taking into account the effects of dust extinction. Although  all our  models were run in halos of similar characteristics, the resulting galaxies present a wide variety of properties, mostly due to the stochasticity to the diversity  of merger histories possible for halos of this type \citep{2005_Bullock_Johnston, cooper2010, Tumlinson2010, AURIGA,2019MNRAS.485.2589M}.

The Auriga simulations have been extensively analyzed in the past, showing that the associated galaxy formation model can generate realistic late-type galaxy models. In addition to their star formation histories, stellar masses, sizes, rotation curves and HI content, previous studies have carefully characterized the effect that their merging activity has in their stellar components. In particular, \citet{2016MNRAS.459L..46M,2019MNRAS.485.2589M} studied in detail the global and radial properties of the stellar halos in the Auriga simulations, showing that they are diverse in their masses and density profiles; mean metallicity and metallicity gradients; ages; and shapes, reflecting the stochasticity inherent in their accretion and merger histories. Furthermore, a comparison with observations of nearby late-type galaxies (mainly from the GHOSTS survey, \citealt{2016MNRAS.457.1419MonaBell, 2017MNRAS.466.1491H}) shows very good agreement between most observed and simulated halo properties. \citet{2018MNRAS.478..548S} studied the present-day satellite luminosity functions of the Auriga halos and found that they are in excellent agreement with those observed in the MW and M31. Furthermore, they also showed that the cumulative satellite mass distribution is  converged for stellar masses $\gtrsim 10^6$ M$_{\odot}$ at the resolution level used in this work. This convergence is reached at masses well below the satellite masses considered in this work. With respect to the internal properties of the Auriga satellites, \citet{2021MNRAS.507.4953G} \citep[see also][]{2019MNRAS.486.4790B} showed that, within the mass/luminosity range considered in this work, the satellite phase-space scaling relations obtained from  the populations of simulated satellites follow reasonably well the observed scaling relations \citep{2012AJ....144....4M}. Moreover, they also showed that the scaling relations are well converged over 3.5 orders of magnitude in mass resolution.

Thus, the Auriga simulation suite represents a suitable simulation set to study the properties of the brightest stellar streams and their progenitors.

\section{The brightest stellar streams}
\label{sec:brightest stellar}

In this section, we describe the procedure followed to identify and quantify the present-day brightest stellar stream in each Auriga model, associated whith either previous or ongoing accretion events. The main steps consist of the generation of different surface brightness (SB) maps, each at a different SB limiting value, for each Auriga model. The maps are then visually inspected, one by one, to determine at what limiting SB the brightest stellar stream in each halo can be identified. Once such stream is identified, the properties of its parent satellite galaxy are extracted and analyzed. We highlight that  in this work we only consider as streams low surface brightness features that are of an accreted origin. Thus, low surface brightness substructures of disc origin, such as galactic feathers, are not considered. Finally, we focus our analysis on satellites with satellite-to-host mass ratio $\lesssim 1/4$ that crossed the viral radius earlier than 1 Gyr ago, thus discarding very recent accretion events.

\begin{figure*}
    \centering
    \includegraphics[scale=0.5,trim={2cm 0cm 5cm 3cm}, clip]{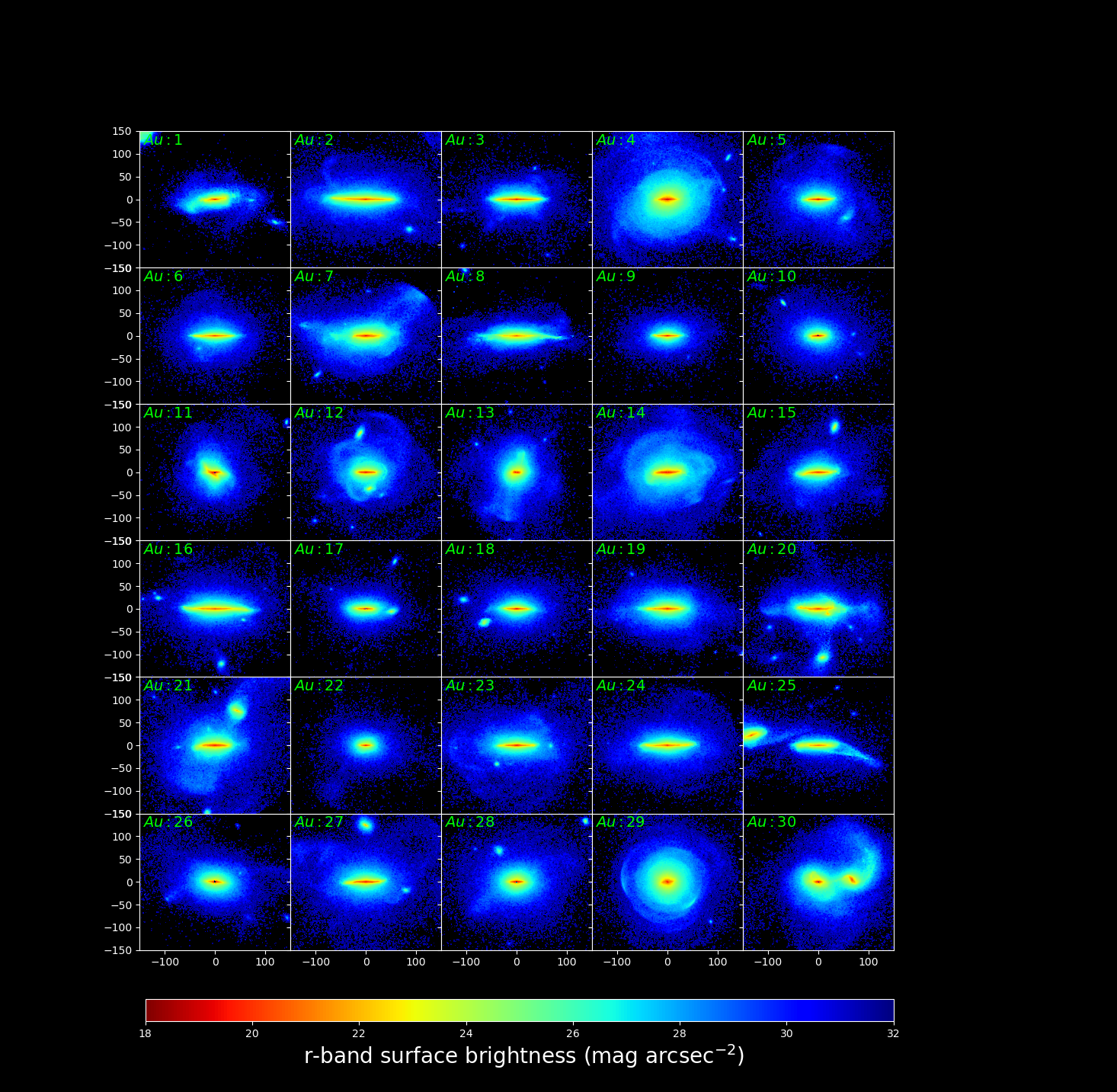}
    \caption{$r$-band surface brightness maps to $\mu_{\rm r}^{\rm lim} = 31$ mag arcsec$^{-2}$ of the Auriga halos at $z=0$, seen edge-on in a square of $300\times 300$ kpc$^2$. The resolution of these maps is $1\times1$ kpc$^2$ per pixel. In most case it is possible to appreciate low surface brightness features that extend well outside the host stellar disc, associated with both previous and ongoing accretion events. }
    \label{fig:mosaico1}
\end{figure*}

\subsection{Surface Brightness Maps}

To generate the  SB-maps, each simulation is first projected onto two planes; i.e. it is rotated such that {\it i)} the angular momentum of the disc is aligned with the z-axis of the reference frame (edge-on) and {\it ii)} the angular momentum of the disc is aligned perpendicularly to the z-axis of the reference frame (face-on). For each projection,  the rotation is done iteratively using only young stellar particles (age $< 5$ Gyr) located within a  cylindrical volume of 10 kpc radius and shrinking height, $h = (10,~5,~2.5)$ kpc. We choose these two particular  disc orientations because they represent the two most extreme configurations to identify low surface brightness features. Whereas on a face-on configuration stellar streams can be hidden by the presence of the much brighter disc out to larger radii (e.g. up to 50 kpc or more), on an edge-on view the disc contamination is minimal beyond 5-10 kpc along the minor axis. We  then center a  $150 \times 150$ kpc two-dimensional grid on top of both projections, using  grid nodes separated by 1 kpc on each direction. Centered on each grid node, we place bins of $2 \times 2$ kpc size, and calculate the total r-band magnitude by integrating over the fluxes of all enclosed stellar particles.  It is worth noting that this gridding of the data, as well as its bin size, mimics the smoothing done in observations to enhance diffuse structure and preserve image resolution \citep[see][]{2018A&A...614A.143M}. As mentioned in Section~\ref{sec:auriga}, the $r$-band luminosity of each galaxy was modelled using \citet{BruzualCharlot} stellar population synthesis models \citep{AURIGA}. We focus on this synthetic band since {\it i)} it is a relatively good tracer of the overall mass distribution of satellites, {\it ii)} it is a typical band employed to observe low surface brightness (LSB) features \citep[e.g.,][]{2013ApJ...765...28A,2019hsax.conf..146M} and {\it iii)} it is less affected by dust extinction than other bluer photometric bands. We note that the brightest streams in the $r$-band are also the brightest in the other simulated photometric bands. If we instead consider, e.g., the $z$ ($B$)-band luminosity, the limiting SB level at which the brightest stream are detected will systematically shift by $\lesssim 1$ magnitude brighter (fainter). Other than that, our overall results remain the same.

Figure \ref{fig:mosaico1} shows the deepest edge-on SB maps obtained from each Auriga model, $\mu_{\rm r}^{\rm lim} = 31$ mag arcsec$^{-2}$. We can see that, in most halos, a clear stellar stream can be found. Note, however, exceptions such as Au 9 and Au10 which do not show any clear LSB feature. As discussed by \citet[][]{2015MNRAS.454.2472H} \citep[see also][]{10.1093/mnras/stz1251}, the shape of the substructures found in these halos is related to the mass and orbital properties of the infalling satellites, such as the impact parameter and the inclination angle with respect to the disc plane. In general, extended stellar streams, such as loops, are typically related to satellites with large impact parameters ($\alpha \geq 10^{\circ}$) and  angular momenta. On the other hand, shell-like structures are associated with the accretion of satellites on nearly radial orbits \citep{cooper2011}. The latter tend to have shorter life times that loops (see also J08). Note as well that debris from satellites can also be found on the plane of the disc. Such streams are the result of either the accretion of  satellites on low inclination infall orbits, or to the tilting of the disc due to the torque exerted by a massive infalling satellite \citep[see e.g.][]{2017MNRAS.465.3446G,2017MNRAS.472.3722G}.

We then generate, for each Auriga model, several SB maps reaching different limiting magnitudes, emulating different observational depths. The limiting SB magnitude, $\mu_{\rm r}^{\rm lim}$, ranges from 22 mag arcsec$^{-2}$ to 31 mag arcsec$^{-2}$, with a step of 1 mag arcsec$^{-2}$. As discussed by \citet{2018A&A...614A.143M}, for SB levels deeper than 28 mag arcsec$^{-2}$ the contamination from Galactic cirrus becomes very significant. We recall that, in this work, we do not account for the effects of the Galactic cirrus. We also do not model internal dust extinction or background subtraction noise which affects detection of LSB features in photometric images. As a result, our analysis  could slightly overestimate the detectability fraction at a given  $\mu_{\rm r}^{\rm lim}$. It is worth noting, however, that we are only focusing on the brightest stream on each model, and that the relatively coarse steps in SB magnitude used to generate the maps, minimize this effect. We will explore the effect of such contamination in a future work.

\begin{figure*}
	\includegraphics[width=\textwidth, trim={4cm 0 9cm 0}, clip]{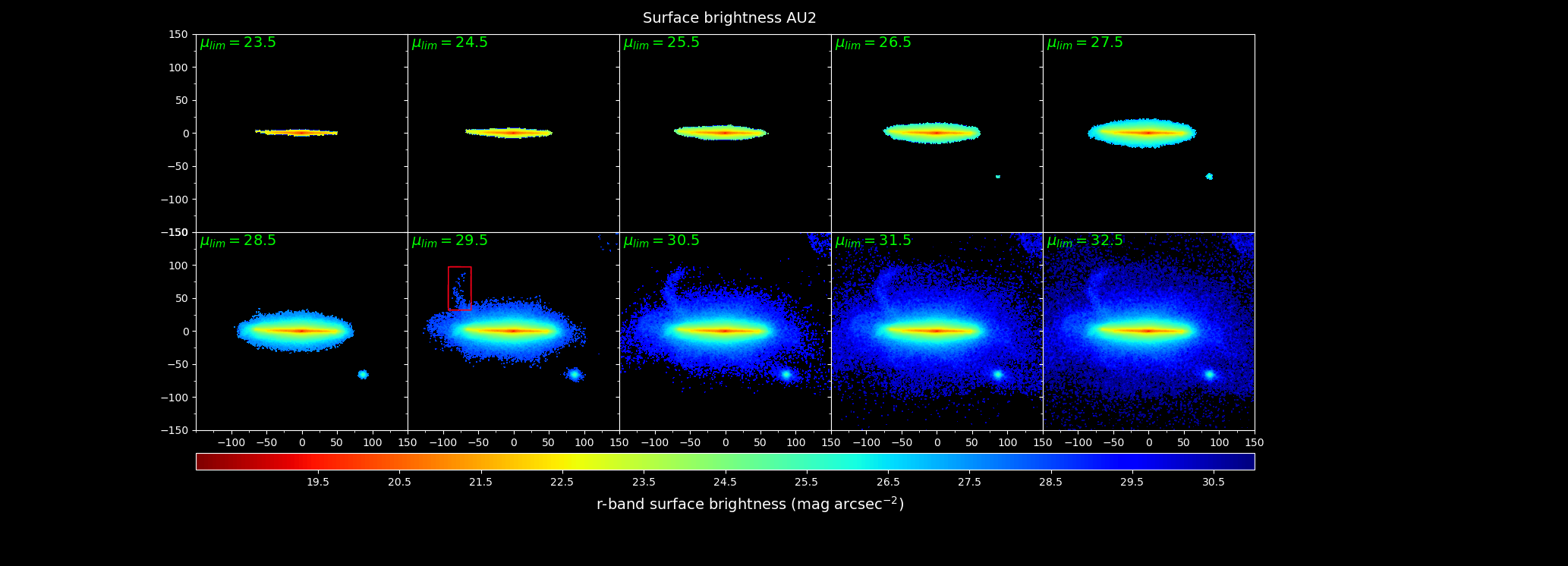}
    \caption{Surface brightness maps as a function of $\mu_{\rm r}^{\rm lim}$, for Au2. The galaxy is shown on its edge-on projection. Note that the brightest stellar stream in this galaxy can be clearly seen for the first time at  $\mu_{\rm r}^{\rm lim} = 29.5$ mag arcsec$^{-2}$. The corresponding stream is highlighted with a red box. The size of each projection is  $300 \times 300$ kpc$^2$. }
    \label{fig:SB_Au2}
\end{figure*}

\begin{figure*}
	\includegraphics[width=\textwidth, trim={4cm 0 9cm 0}, clip]{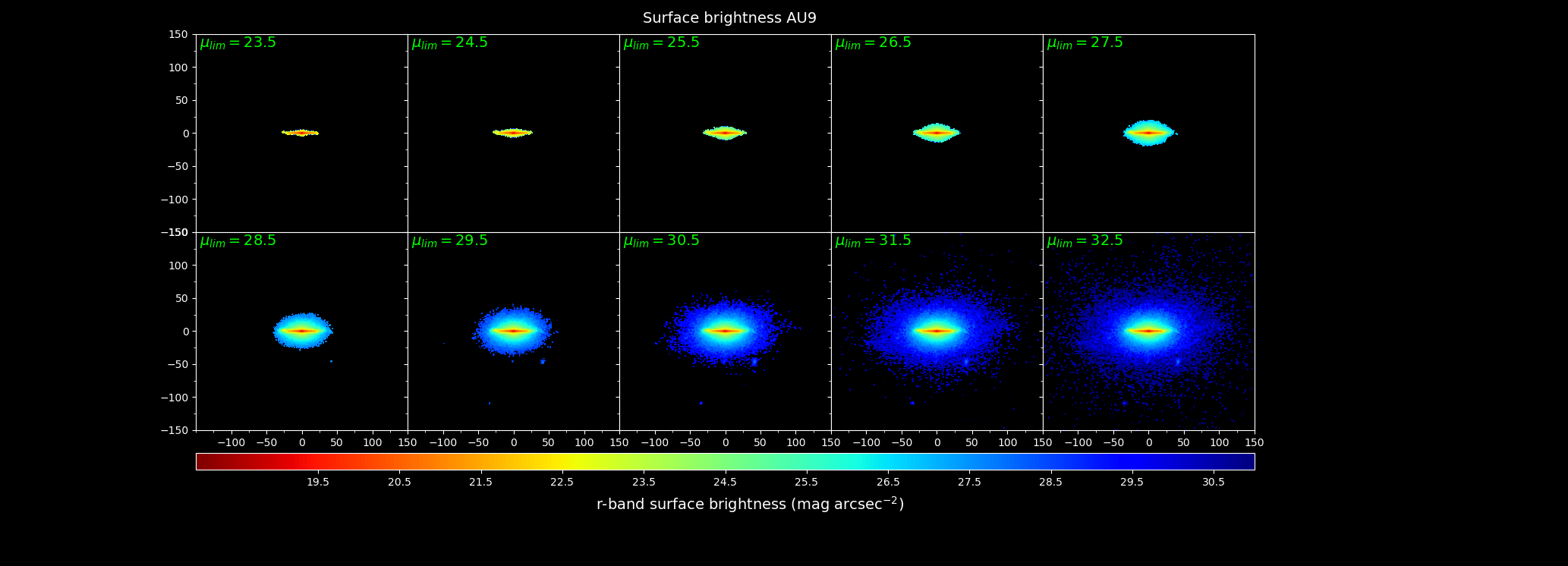}
    \caption{Surface brightness maps as a function of $\mu_{\rm r}^{\rm lim}$, for Au9. The galaxy is shown on its edge-on projection. Note that no clear stellar stream is found at any surface brightness. The size of each projection is  $300 \times 300$ kpc$^2$. }
    \label{fig:SB_Au9}
\end{figure*}

\begin{figure*}
    \centering
    \includegraphics[scale=0.5,trim={2cm 0cm 5cm 3cm}, clip]{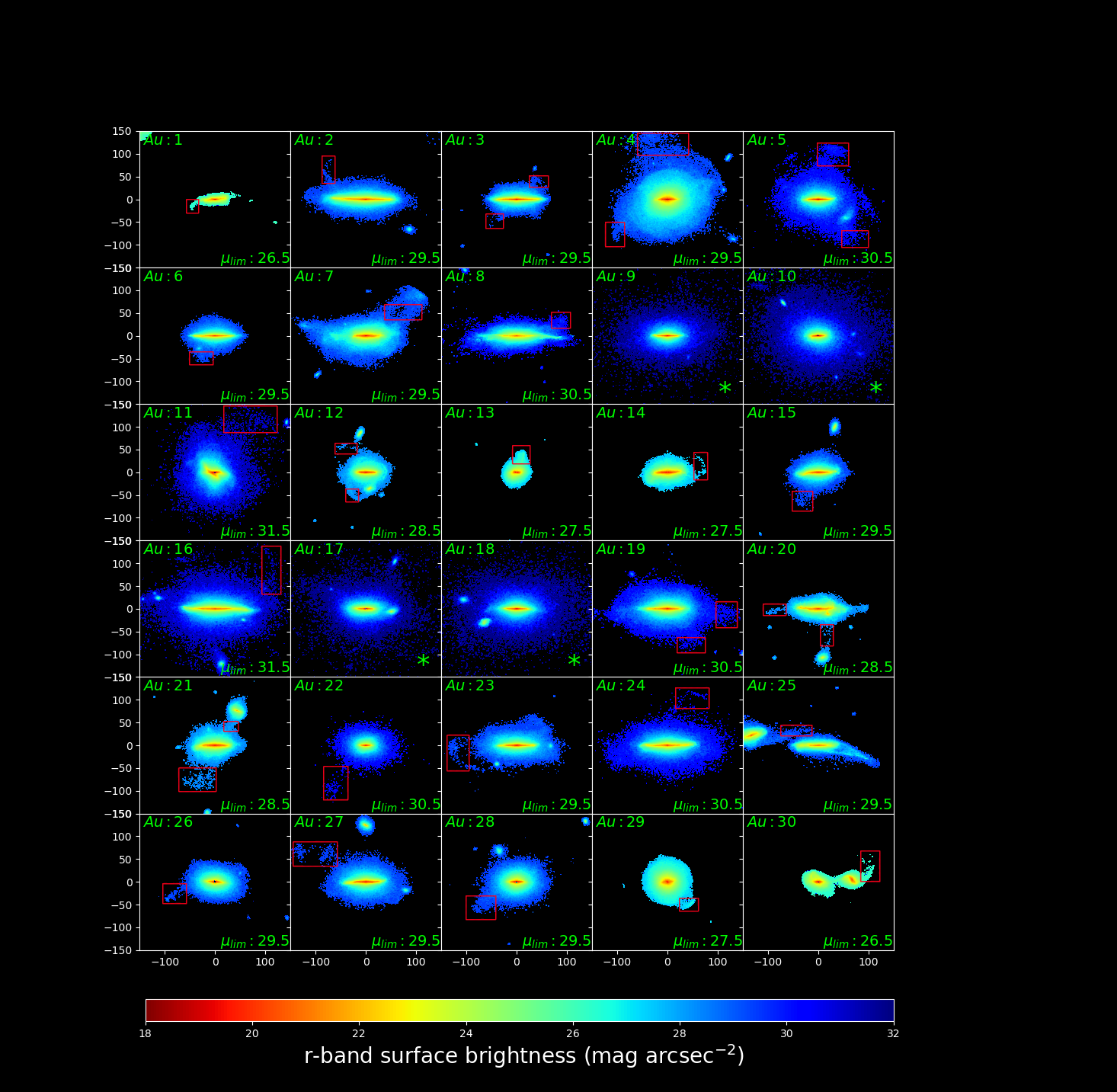}
    \caption{As in Figure\ref{fig:mosaico1} but now selecting, for each halo, the SB map at the  $\mu_{\rm r}^{\rm lim}$ value where the brightest stellar stream is first identified. The red boxes highlight the brightest stellar streams in each halo. Some show two boxes. This indicates that two different streams were identified at the same surface brightness. Note that some halos (Au9, Au10, Au17, Au18) do not show identifiable stellar streams. Those halos are highlighted with an * symbol.}
    \label{fig:mosaico2}
\end{figure*}

In Figure \ref{fig:SB_Au2} we show an example of the results obtained with this procedure. The figure shows SB maps of the Au2 model, displayed on an edge-on projection. The different panels show the resulting SB maps obtained for different values of $\mu_{\rm r}^{\rm lim}$. As expected, shallow $\mu_{\rm r}$ maps ($\mu_{\rm r}^{\rm lim} \lesssim 24$ mag arcsec$^{-2}$) only reveal the  presence of the bright stellar disc and, if present, of the brightest satellite galaxies. In general, at $\mu_{\rm r}^{\rm lim} \sim 25$ mag arcsec$^{-2}$ we reach the outer edges of all stellar discs.  Deeper maps start to reveal a more extended, relatively flat stellar distribution associated with the inner stellar halo, mainly dominated by an in-situ component \citep{2019MNRAS.485.2589M}, as well as the faint and extended stellar halo. These deeper maps allow us to detect, in many cases, low surface brightness substructures mostly associated with stellar streams from ongoing or previous accretion events. In the following section we describe the procedure applied to identify the brightest stellar stream in each simulated halo using these SB maps.

\begin{figure}

    \includegraphics[width=\columnwidth]{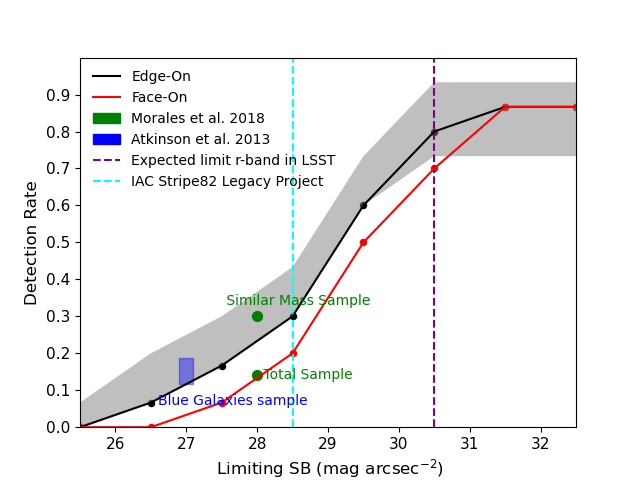}
    \caption{Cumulative fraction of the number of halos with detected stellar streams as a function of $\mu_{\rm r}^{\rm lim}$. The black and red lines show the results obtained when galaxies are oriented edge-on and face-on, respectively. The grey shaded area shows the range obtained from the five independent stellar streams visual identifications. The blue shaded area shows the fraction of blue galaxies sample with stellar streams, detected within the second and third highest confidence level by \citet{2013ApJ...765...28A} and green circles show the fraction of galaxies with identifiable stellar streams reported by \citet{2018A&A...614A.143M}. The dashed lines indicate the limits in surface brightness of different surveys, in cyan the IAC Stripe82 Legacy Project, in purple the limit expected for the ten years of co-added Vera Rubin Observatory’s LSST images.} 
    
    \label{fig:cumulative}
\end{figure}

\begin{table}
	\centering
	\caption{ Properties of the brightest stream satellite progenitors. The columns are (1) the Auriga galaxy halo number; (2) the satellite lookback infall time, or time when the satellite first crossed their host virial radius; (3) the logarithm of the total mass of the satellite at infall time: $\log(M_{\rm tot}/M_{\odot}$); (4) the logarithm of the stellar mass of the satellite at infall time: $\log(M_{\rm \star}/M_{\odot}$); (5) flag that indicates whether satellites survive at $z=0$ or not; (6) satellite's ranking, $\#_{\rm prog}$, based on the stellar mass contributed to the $z=0$ host stellar halo mass; (7) the logarithm of the total mass of the host galaxy at $z=0$, $\log(M_{\rm tot}^{\rm Host}/M_{\odot}$). Galaxies that show multiple streams, above the surface brightness threshold, associated with different satellites are highlighted with an {\it $(a)$}.}
	\label{tab:Satellite properties}
	\begin{tabular}{ccccccc}
		\hline
		Au & $t_{\rm infall}$(Gyr) & $M_{\rm tot}^{\rm Sat}$      & $M_{\star}^{\rm Sat}$ & Surv.  &  $\#_{\rm prog}$ & $M_{\rm tot}^{\rm Host}$ \\

		\hline
		1      & 4.56 & 10.93 & 9.57  & NO   & 1 & 11.98  \\
		2      & 4.24 & 10.30 & 8.53  & YES  & 5 & 12.3   \\
		3      & 9.42 & 10.34 & 9.13  & NO   & 2 & 12.18  \\
		4$^a$  & 3.11 & 11.49 & 10.37 & NO   & 1 & 12.17  \\
		       & 6.32 & 10.70 & 9.03  & NO   & 3 &        \\
		5$^a$  & 8.49 & 10.27 & 8.46  & NO   & 3 & 12.1   \\
		       & 6.78 & 10.29 & 8.45  & YES  & 4 &        \\
		6      & 8.96 & 10.43 & 8.70  & NO   & 1 & 12.04  \\
		7      & 2.79 & 11.04 & 9.80  & NO   & 1 & 12.07 \\
		8      & 8.03 & 10.84 & 9.38  & NO   & 1 & 12.04  \\
		9      & ---- & ----- & ----  & ---  & --& 12.04  \\
		10     & ---- & ----- & ----  & ---  & --& 12.04  \\
		11     & 0.82 & 11.47 & 10.51 & YES  & --& 12.23  \\
		12     & 5.68 & 10.67 & 9.28  & NO   & 3 & 12.06  \\
		13     & 7.25 & 10.82 & 9.33  & NO   & 1 & 12.1  \\
		14     & 7.25 & 10.92 & 9.45  & NO   & 6 & 12.24  \\
		15     & 5.52 & 10.51 & 9.10  & YES  & 6 & 12.1  \\
		16     & 8.65 & 9.73  & 7.68  & YES  & 10& 12.19 \\
		17     & ---- & ----- & ----  & ---  & --& 12.04 \\
		18     & ---- & ----- & ----  & ---  & --& 12.11\\		
		19$^a$ & 7.10 & 10.61 & 9.14  & YES  & 2 & 12.1 \\
		       & 5.99 & 10.41 & 8.89  & NO   & 3 &      \\
		20$^a$ & 5.99 & 11.28 & 9.98  & NO   & 1 & 12.11  \\
		       & 5.04 & 10.49 & 9.05  & YES  & 4 &   \\
	    21$^a$ & 6.47 & 10.80 & 9.16  & NO   & 3 & 12.18 \\
		       & 4.24 & 11.00 & 9.65  & YES  & 4   \\
		22     & 5.52 & 10.23 & 8.44  & NO   & 1 & 11.99 \\
		23     & 6.63 & 10.38 & 8.70  & YES  & 3 & 12.22 \\
		24     & 9.88 & 10.33 & 8.48  & NO   & 4 & 12.19  \\
		25     & 1.65 & 11.32 & 10.21 & YES  & 1 & 12.1 \\
		26     & 8.96 & 10.47 & 8.86  & NO   & 2 & 12.22  \\
		27     & 6.94 & 10.69 & 8.93  & NO   & 3 & 12.26 \\
		28     & 7.25 & 9.94  & 8.73  & YES  & 4 & 12.23 \\
		29     & 5.04 & 11.56 & 10.48 & NO   & 1 & 12.21  \\
		30     & 3.60 & 11.15 & 9.92  & YES  & 1 & 12.06 \\
		\hline
	
	\end{tabular}

\end{table}

\subsection{Identification of the brightest stellar streams}
\label{sec:identify}

Our main goal in this work is to characterize what information can be extracted from the brightest stellar streams, of accretion origin, with respect to the recent accretion history of a galaxy. Thus, the first step is to identify such streams in the stellar halos of each Auriga model, and their different projections. The automatic detection of stellar streams is a challenging task. These LSB features can show a wide variety of morphologies and are typically very extended, sampling galactic regions with very different SB levels. Methods to achieve this have been previously proposed, especially when dealing with large number of observations where visual identification is neither scalable nor feasible \citep[see e.g.][]{2018ApJ...866..103K}. In our work, we have 30 galactic models and, thus, visual inspection of SB maps to identify the brightest stream on each halo can be reliably applied. We also note that visual inspection is the preferred method to identify LSB features in most observational works and, in particular, those that we discuss and compare against in Section 3 such as \citet[][]{2013ApJ...765...28A} and \citet[][]{2018A&A...614A.143M}.

To account for human bias on the visual inspection of SB maps, we proceed as follows. First, among five co-authors, we distributed several SB maps with different $\mu_{\rm r}^{\rm lim}$ for each galactic model. Examples of these maps are shown in Fig.~\ref{fig:SB_Au2} and \ref{fig:SB_Au9}, where each panel shows the result of reaching a progressively deeper $\mu_{\rm r}^{\rm lim}$. These maps include all stellar particles within the given area, independently of whether they were born in situ, i.e. within the potential well of the main host, or in satellite galaxies, i.e. accreted particles. On each model we start searching for LSB features on the shallowest SB map. Each person was requested to identify LSB features independently of their projected morphology. As a result, the brightest LSB features in our analysis could be associated with tidal streams, shells or plume-like structures. Note that  LSB features that are associated with main galactic hosts, such as galactic feather or bound satellite galaxies, are not accounted for since we focus on features of an accreted origin. If no LSB features are identified at a given $\mu_{\rm r}^{\rm lim}$, we proceed to the next deeper map. This is iteratively performed until the brightest stream can be first identified. Once the stream is observed, the corresponding value of $\mu_{\rm r}^{\rm lim}$ is stored on a list. In the example shown in Fig.~\ref{fig:SB_Au2}, a clear tidal stream,  highlighted with a red box, is first detected at $\mu_{\rm r}^{\rm lim} = 29.5$ mag arcsec$^{-2}$. Note however that in some models, such as Au9, shown in Fig.~\ref{fig:SB_Au9}, it was not possible to identify a well defined LSB feature at any $\mu_{\rm r}^{\rm lim}$.

Secondly, these five co-authors shared their independent identification list to reach a  consensus. A unique and final identification list is obtained from this procedure. Those cases where discrepancies in the identifications were found were individually discussed. A common consensus was reached after different checks, such as whether the identified features were not a low surface brightness bound satellite. Once a final list is obtained, the origin of these LSB features is asserted based on the birth location of the stellar particles that composed them. In those cases where the identified substructure is associated with debris from the main galactic host, we discarded it and proceeded to identify the following brightest LSB feature following the procedure just described.

The result of this identification process is show in Figure~\ref{fig:mosaico2}. Each panel shows the shallowest SB map in which the brightest  stream has been clearly identified. The corresponding limiting SB is listed on the legend; the cases where no substructure is present have the SB marked with an * symbol. The brightest stream in each halo is highlighted with a red box. In some cases, such as Au12, Au20, Au21 and Au25, the brightest stream can be directly linked to the brightest satellite in the field. This shows that those satellites have been orbiting their host for a few Gyrs and are currently undergoing disruption. Such streams can be observed in relatively shallow maps, reaching $\mu_{\rm r}^{\rm lim} \lesssim 27$ mag arcsec$^{-2}$. In many other cases, such as Au5, Au11, Au16 and Au22, it is necessary to reach much deeper SB levels to  identify the brightest stream. These are clear examples of how low surface brightness substructures can go undetected without very deep observations. Note that, in those cases, the progenitor satellites can no longer be identified.  We also find halos, such as Au18, where the brightest satellite does not show any tidal feature indicating that they have been very recently accreted onto their host.

\subsection{Brightest stellar stream quantification}

\begin{figure*}
	\includegraphics[width=\textwidth, trim={1cm 0 2cm 0}, clip]{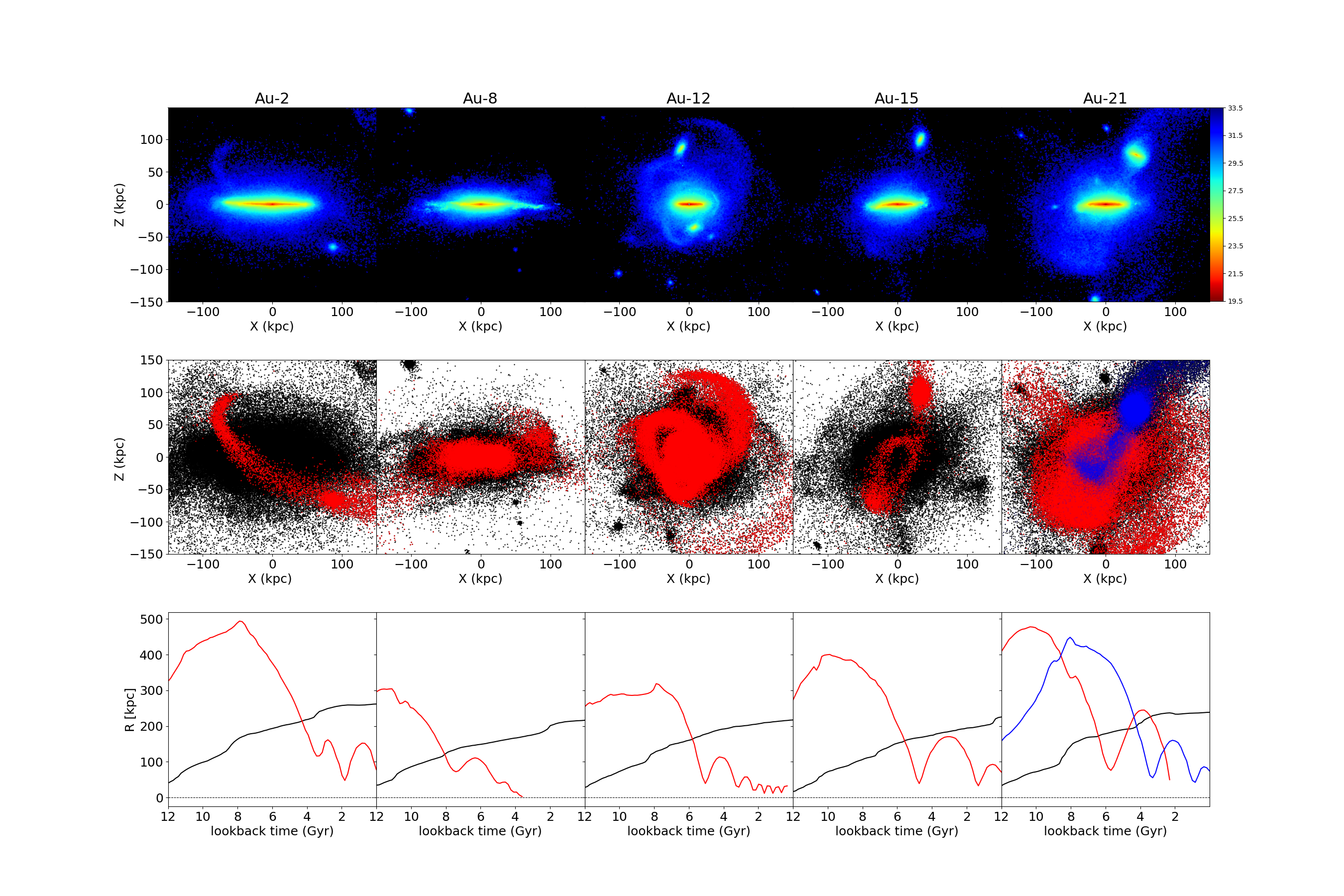}
    \caption{Top panels: Surface Brightness maps of five different Auriga galaxies, shown edge-on in a $300\times300$ {\rm kpc$^2$} area. Middle panels: the black dots show the stellar particle distributions of these halos.  Red and blue dots show the stellar particles associated with the brightest stream progenitors on each halo. The different colors in  Au21 correspond to particles from two different satellites that contributed with streams at the same $\mu_{\rm r}^{\rm lim}$.
    Bottom panels: The black line shows the time evolution of the host virial radius. The red and blue lines show the time evolution of the galactocentric distance of the brightest stream progenitors. The dashed line indicates $R=0$ kpc. }
    \label{fig:triple}
\end{figure*}

In Figure~\ref{fig:cumulative} we summarize the results of the brightest stream identification process. We first focus on the results obtained from the edge-on projection. The black line shows the normalized cumulative function of the number of models with identifiable stellar streams as a function of the limiting SB level, $\mu_{\rm r}^{\rm lim}$, for the edge-on projections. The grey shaded area indicates the full range obtained from the five independent stellar streams identifications. To define the lower and upper bound of this area, we assigned to each model the maximum and minimum $\mu_{\rm r}^{\rm lim}$ value reported by the five identifiers. None of the Auriga models shows stellar streams in SB maps with $\mu_{\rm r}^{\rm lim} \leq 25.5$ mag arcsec$^{-2}$. However, we find that at  $\mu_{\rm r}^{\rm lim} \approx 28.5$ mag arcsec$^{-2}$, where contamination from Galactic Cirrus starts to become significant, $\sim 30$ percent of the models already show a detection. The cumulative function shows a steep increase in the fraction of galaxies with stream detection  at this $\mu_{\rm r}^{\rm lim}$ value, and then flattens again beyond 30.5 mag arcsec$^{-2}$. Another interesting result is that, even at a very low SB limit of 31 mag arcsec$^{-2}$, 13 percent of our models show no detectable stream. The cyan dashed line in Fig.~\ref{fig:cumulative} shows the deepest SB achieved by the SDSS through its IAC Stripe 82 Legacy Survey \citep{2016MNRAS.456.1359F}, which is at $\mu_{\rm r}^{\rm lim} \approx 28.5$ mag arcsec$^{-2}$ in the r-band. Note that, as mentioned above, at this SB limiting magnitude we are able to detect the brightest streams on $\approx 30$  percent of late-type galaxies. For a more complete census of these LSB features it is thus critical to obtain  deeper surveys such as what will be provided by the LSST from the Vera-Ruby observatory. The final 10 years LSST catalogue is expected to provide images over a wide area (18,000 square degrees) reaching SB magnitudes of 30.5 mag arcsec$^{-2}$ in the r-band \citep{2018arXiv181204897L}. This is highlighted in Fig.~\ref{fig:cumulative} with the purple dashed line.

We note that two models show as brightest LSB features material directly linked to the host stellar discs. These two cases are Au11 and Au25 and both have suffered a recent and strong tidal interaction. These substructures, best known as galactic feathers, are not very common in our sample, showing in only 5 per cent of the objects. Since feathers are associated with the host pre-existing disc, they are typically very bright. In the two cases found here, they can be clearly already seen at SB brighter than $\sim$ 27 mag/arcsec$^2$. Substructure from the perturbing satellites is typically also present, but usually is observed at fainter magnitudes. Feathers are not included in the cumulative function shown in Fig.~\ref{fig:cumulative}. Instead, we focus on the brightest LSB features associated to satellites, which arise at a lower SB level.

It is interesting  to compare these results with those obtained from observational samples. For example, \citet{2013ApJ...765...28A}, using observations from the wide-field component of
the Canada–France–Hawaii Telescope Legacy Survey, generated a sample of 1781 luminous ($M_{r^\prime }<-19.3$ mag) galaxies in the magnitude range 15.5 mag $< r^{\prime } < 17$ mag, and in the redshift range $0.04 < z < 0.2$. The sample reaches a limiting surface brightness in the $r^\prime$-band of $\sim 27$ mag arcsec$^{-2}$ and  it was visually inspected to detect LSB features. Their analysis show that $12$ percent of the galaxies in their sample present clear tidal features at the highest confidence level, but the fraction rises to about $26$ percent if systems with marginal detection are included.  Note however, that the sample studied by \citet{2013ApJ...765...28A} includes galaxies lying in both the red sequence and the blue cloud. Furthermore, they find that the fraction of galaxies with detected streams is a strong function of the rest-frame color and stellar mass, and that red galaxies are twice as likely to show tidal features than blue galaxies.   In our work, we are biased towards very bright late-type galactic models, with $-23 <M_{r^\prime }<-20$ mag, with median $M_{r^\prime } \approx -22$ mag. Thus, our models fall within the blue cloud (see Fig. 20 in \citealt{AURIGA}). The blue cloud subsample by \citet{2013ApJ...765...28A} shows tidal detections in 12 to 18 percent of their galaxies, depending on the confidence level detection considered, which is shown in Fig.~\ref{fig:cumulative} as a blue shaded area. This percentage is in good agreement with our results. Around their value of $\mu_{\rm r}^{\rm lim}= 27$ mag arcsec$^{-2}$, we find that $\sim 15$ percent of our models show low surface brightness features.

Similar results were obtained by \citet{2018A&A...614A.143M} using a sample of a post-processed  Sloan Digital Sky Survey (SDSS) images, optimized for the detection of stellar structures with low surface brightness around a volume-limited sample of nearby galaxies. Their final sample consists of images of 297 galaxies with stellar masses similar to that of the Milky Way, which are visually inspected by the authors to detect LSB features. The images sampled reach a Gaussian distributed  $\mu_{\rm r}^{\rm lim}$ of mean $\approx 28.1$  mag arcsec$^{-2}$ and $\sigma \approx 0.26$ mag arcsec$^{-2}$. Within those limiting SB, they find a detection of stellar substructure in $14$ percent of the observed galaxies, whereas we find stellar streams in almost 25 percent of the Auriga models (see Fig.~\ref{fig:cumulative}). 
An important difference between the Morales sample of observed galaxies and our sample of models, which most likely accounts for this mismatch, is the stellar mass distribution of galaxies. The observed sample analyzed by \citet{2018A&A...614A.143M} has a mean stellar mass of log($M_*/M_{\odot}) = 10.37$ whereas the mean stellar mass of the Auriga models is log($M_*/M_{\odot}) = 10.82$. We note also that all our models are more massive that the median mass of the observed sample. This is rather significant, especially considering that the observed stellar substructure detections increase significantly for larger stellar masses  of the host galaxy, with a detection rate of about $\sim 30$ percent for stellar masses larger than log($M_*/M_{\odot}) = 10.82$ (See Fig. 7 of \citealt{2018A&A...614A.143M}).

There are several other differences between the analysis presented in this work and those based on observational samples of galaxies which should be kept in mind while performing this comparison. It is worth recalling that, in our models, {\it i)} we are not accounting for dust extinction and background noise, which may erase the signature of faint stellar substructures, {\it ii)} we have a much smaller sample of galaxies, and {\it iii)} we count as a detection the very first time we see signs of a stream. In reality it is likely that slightly deeper observations would be required in some of these cases to detect streams due to the above-mentioned observational effects. It is also worth recalling that in both observational studies, \citet{2013ApJ...765...28A} and \citet{2018A&A...614A.143M}, stellar streams were also identified by visual inspection of the corresponding images, as also done in this work.

\begin{figure}

    \includegraphics[width=\columnwidth]{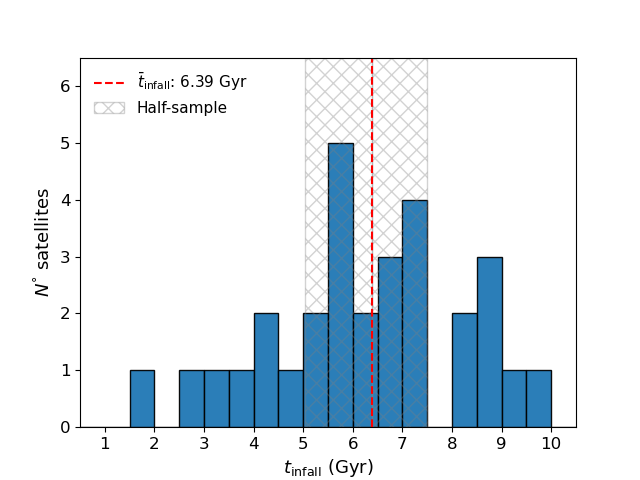}
    \caption{Lookbak infall time distribution of all brightest stream progenitors. The distribution has a median value of 6.39 Gyr.  The cross-hatched area, centred on the median, encloses $50\%$ of the sample and ranges from 5 to 7.5 Gyr. }
    \label{fig:hist_time}
\end{figure}

Our results discussed so far are based on edge-on projections of the galactic models. As previously mentioned, this is bound to enhance the detection of streams when compared with a sample of galaxies distributed at random inclinations. To explore this effect we show  in Figure~\ref{fig:cumulative}, with a red line, the cumulative fraction of models with detected streams as a function of $\mu_{\rm r}^{\rm lim}$ when projected face-on.  Indeed we find that this fraction decreases by $\approx 10\%$ within the relevant SB magnitude range.

To summarize, it is expected that the fraction of galaxies with detected streams in this work should be somewhat larger than that found in observations, which are likely to represent a lower detection limit. Nevertheless, we find a reasonably good agreement between our detection rates and those from the observed samples. In a follow-up work we will include  observational effects and biases to our models in order to perform a more quantitative and fair comparison with observational results. 


\section{Properties of satellite progenitors}
\label{sec:properties}

In the previous section we have identified and quantified the brightest stellar streams in each Auriga halo by inspecting SB maps at different $\mu_{\rm r}^{\rm lim}$ and  projections. The goal of this  section is to identify and characterize the main properties of the satellite progenitors of these detected low surface brightness features, in particular their infall times and masses. 
The progenitor satellite of a given stream is identified by searching among the satellites that contributed the largest number of particles to a small area surrounding the brightest stream. These areas for each Auriga model are highlighted with a red square in Figure \ref{fig:mosaico2}. This procedure  is summarized in Figure~\ref{fig:triple}. The top panels shows, as an example, the SB maps of five Auriga models reaching $\mu_{\rm r}^{\rm lim} = 31$ mag arcsec$^{-2}$. The middle panels show stellar particle scatter plots of the same halos. With red dots we highlight the stellar particles that belong to the brightest stream progenitor detected in each halo. 
Some models such as Au21 (rightmost panel) show  more than one stream at the corresponding  $\mu_{\rm r}^{\rm lim}$ where substructures were first revealed (see also Fig.~\ref{fig:mosaico2}). In those cases where the two streams are associated to two different progenitors, both satellites are highlighted with red and blue dots. Note that, while some satellite progenitors can still be identified, i.e. are surviving satellites (e.g., Au2 and Au21-blue dots), in other cases, they have been fully disrupted (e.g. Au11 and Au21-red dots). We will further explore this in what follows.

\subsection{Progenitor infall times}
\label{sec:time}

Once a progenitor satellite is identified, we proceed to trace it back in time. This is done by following the corresponding merger trees. Note that in Auriga every satellite is assigned a unique identification number. This ID is assigned to every stellar particle born within the potential well of the corresponding satellite. As a result, it is possible to track a satellite's stellar particles even after its full disruption. The bottom panels of Figure ~\ref{fig:triple} show with red and blue lines the evolution of the satellite galactocentric distance, $R_{\rm sat}$, as a function of lookback time.  
For comparison, we also show with a black line the time evolution of the host virial radius, $R_{\rm vir}$.

\begin{figure}      
	\includegraphics[width=85mm,clip]{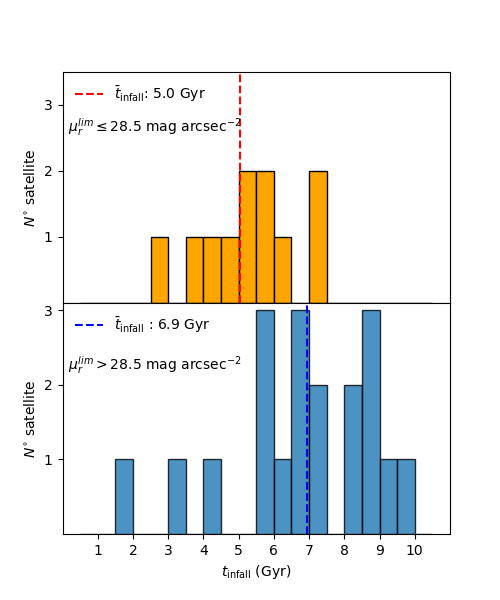}
      \caption{Top panel: distribution of lookback infall times for the BSPs of streams first identified in SB maps with $\mu_{r}^{\rm lim} \leq 28.5$ mag arcsec$^{-2}$. The red line shows the median $t_{\rm infall} = 5$ Gyr. Bottom panel: same as top panel but for the BSPs of brightest streams first identified in SB maps with $\mu_{r}^{\rm lim} > 28.5$ mag arcsec$^{-2}$. The blue line shows the median $t_{\rm infall} = 6.9$ Gyr. }
    \label{fig:ct_mag}
\end{figure}
\begin{figure}
    \includegraphics[width=\columnwidth]{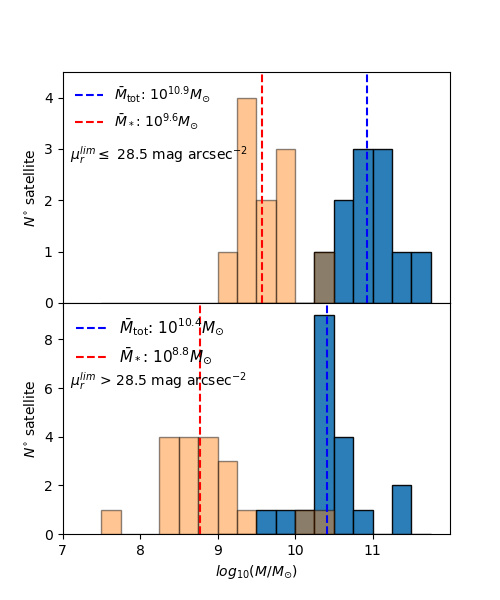}
    \caption{The histograms show the total mass and stellar mass distributions, in blue and orange respectively, for all brightest stream progenitors. Similar to Fig. \ref{fig:ct_mag} the top panel shows the sample for the BSPs of brightest streams first identified at $\mu^{\rm lim}_{r} \leq 28.5$mag arcsec$^{-2}$ and the bottom panel shows the sample for the BSPs of the brightest streams first identified at  $\mu^{\rm lim}_{r} > 28.5$mag arcsec$^{-2}$. The dashed lines indicate the median infall mass for each distribution.}
    \label{fig:dist_mass}
\end{figure}

To estimate the brightest stream progenitor (BSP) lookback infall time, $t_{\rm infall}$, which is the time at which the satellite first crosses the host virial radius, we search for the snapshot where the first minimum of $|R_{\rm sat} - R_{\rm vir}|$ is reached. The $t_{\rm infall}$ associated with those snapshots are listed in Table \ref{tab:Satellite properties}.  In Figure \ref{fig:hist_time} we show the overall  $t_{\rm infall}$ distribution. Interestingly, we find that the satellites that give rise to the brightest stream in each simulation are accreted in a very wide range of times, with $t_{\rm infall}$ values as high and low as 10 and 1.6 Gyr, respectively and a median value of $t_{\rm infall} = 6.39$ Gyr. It is worth highlighting that $50\%$ of the BSPs were accreted within the  range 5 Gyr $\lesssim t_{\rm infall} \lesssim 7.5$ Gyr, as shown by the striped box area. 
As a result, only  $25\%$ of the BSPs corresponds to a very recent accretion event, with $t_{\rm infall} < 5$ Gyr (e.g., Au11). Conversely, the BSPs in $25\%$ of the cases are related to satellites that were accreted as early as 8 to 10 Gyr ago (e.g., Au26).

\begin{figure*}
    \centering
    \includegraphics[width=\textwidth,trim={3cm 0 3cm 0}, clip]{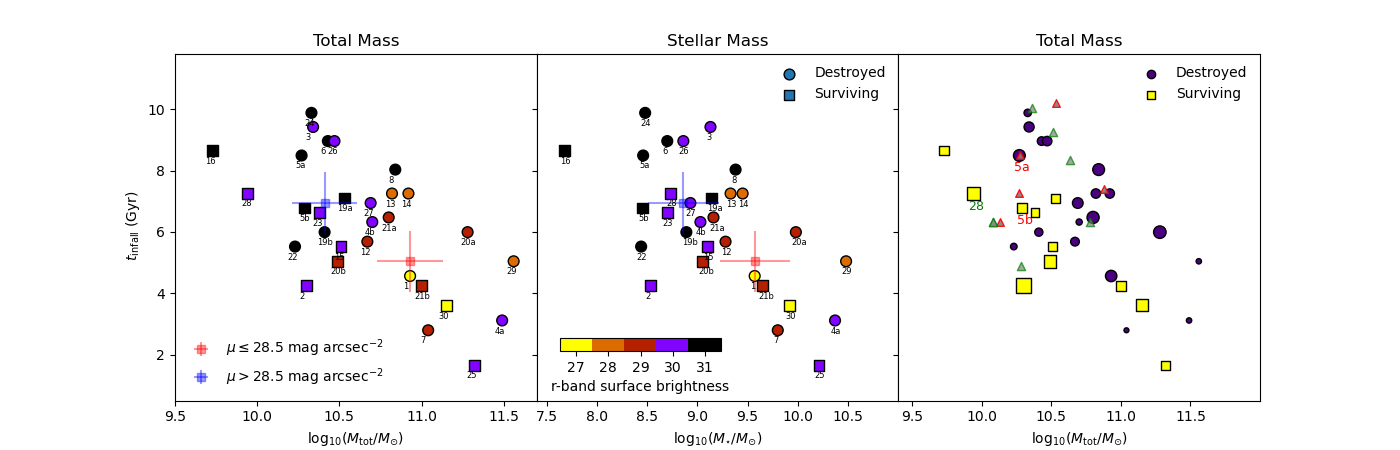}
    \caption{Left panel: distribution of brightest stream progenitor (BSP) lookback infall time as a function of their total mass at infall time. The color coding indicates the $\mu_{\rm r}^{\rm lim}$ at which the brightest stream was first identified. Circles and squares indicate destroyed and surviving BSPs, respectively. The light red and blue crosses show the median $(t_{\rm infall}, M_{\rm tot})$ values for the population of BSPs with stellar streams first identified at $\mu_{\rm r}^{\rm lim} \leq 28.5$ mag arcsec$^{-2}$ and $ > 28.5$ mag arcsec$^{-2}$, respectively. Middle panel: as the left panel, but for $(t_{\rm infall}, M_{*})$ distribution. Right panel: as in the left panel but now symbols are sized according to their first pericentric distance (larger symbols indicate greater distances) and color coded according to whether the BSP survived to the present-day or not. The red and green triangles show the five most massive satellites accreted by halos Au5 and Au28, respectively. This shows that the BSP is not necessarily the most massive nor the most recent accretion event}.
    \label{fig:time_mass}
\end{figure*}

The fifth column of Table~\ref{tab:Satellite properties} indicates whether the BSPs have survived to the present-day or not. The surviving BSPs are those satellites that are still identified by the Subfind routine, at z=0, as an independent subhalo from the main host halo. Interestingly  $\sim 34\%$ of the BSPs can still be identified at $z=0$ as an independent subhalo. The bottom panel of Figure \ref{fig:triple} seems to suggest that, typically, surviving BSPs have crossed $R_{\rm vir}$ at later times than their disrupted counterparts (see e.g. the blue and red line in the last panel of Figure \ref{fig:triple} for an example of a surviving and destroyed satellite, respectively). Indeed, we find that the mean $t_{\rm infall}$ for  the surviving and disrupted BSPs are 5.1 Gyr and 6.7 Gyr, respectively. In this context is worth  taking into account the results of \citet[][hereafter F20]{10.1093/mnras/staa2221}, who analyzed the Auriga simulations to examine the build-up of the MW's stellar halo. Their analysis focused on the comparison between the properties of the surviving and destroyed dwarf galaxies that are accreted by these halos over cosmic time. However, they did not explore the correlation between these accretion events and the low surface brightness substructure left behind. F20 showed that, on average, destroyed dwarfs have  early infall times, $t_{\rm infall} \gtrsim 7.5$ Gyr, whereas the majority of dwarfs accreted at $t_{\rm infall} < 4$ Gyr survive to the present day, in agreement with what we find for the BSPs. Moreover, they also find a dependence between survivability, $t_{\rm infall}$, and satellite mass at infall.  For surviving satellites, the typical $t_{\rm infall}$ are $\sim 8$ Gyr and $\sim 4$ Gyr for satellites with infall stellar masses of $10^6~M_{\odot}$ and $10^9~M_{\odot}$, respectively. Instead, for disrupted satellites, the infall times are $\sim 11.5$ and $\sim 9$ Gyr for infall stellar masses of $10^6~M_{\odot}$ and $10^9~M_{\odot}$, respectively.
Interestingly, one could naively expect the brightest LSB feature to be associated with the most massive accretion event a galaxy has recently experienced. However, as we will discuss later in Section~\ref{sec:mass}, the correlation between satellite infall mass and survivability rate discussed in F20, translates into BSPs not necessarily being the most significant mass contributors to the present-day accreted stellar halos mass distribution.

We now examine the relation between BSPs infall times  and the $\mu_{\rm r}^{\rm lim}$ at wich the brightest streams are first identified. 
For this analysis we divide the sample of BSPs into two subsamples based on the limiting SB at which the stream was first identified: fainter or brighter than 28.5 mag arcsec$^{-2}$. The  division at this limiting SB is motivated by the following.  First, it is at about this SB magnitude where contamination from Galactic cirrus becomes very significant.  Second, and related to the first  point, current large observational surveys that have tried to quantify LSB features on nearby galaxies have not reached deeper (in general) than  $\mu_{\rm r}^{\rm lim}$ =  28.5 mag arcsec$^{-2}$. Future surveys such as those provided by the LSST, will reach much fainter SB levels and we want to show what information will be possible to obtain from these fainter features.
The top and bottom panels of Figure~\ref{fig:ct_mag} show the $t_{\rm infall}$ distribution for the BSPs which streams were first detected at values of  $\mu_{\rm r}^{\rm lim} \leq 28.5$ mag arcsec$^{-2}$ and $\mu_{\rm r}^{\rm lim} > 28.5$ mag arcsec$^{-2}$, respectively. These two subsamples present different distributions. Brightest stellar streams first detected on SB maps at  $\mu_{\rm r}^{\rm lim} < 28.5$ mag arcsec$^{-2}$  represent accretion events that typically took place 5 Gyr ago, while those streams found at $\mu_{\rm r}^{\rm lim} > 28.5$ mag arcsec$^{-2}$ are, on average, related to accretion events that took place about 7 Gyr ago. Note, however, that both distributions show significant dispersion in $t_{\rm infall}$. For this calculation we have removed Au11 whose BSP has just been accreted and, thus, presents a very faint shell-like substructure that has not yet had the time to develop.

The relation between brightest stream brightness and the progenitor infall times is not surprising. If, for simplicity, we assume that the BSP infall mass distribution is similar for both samples, streams that have more time to phase-mix should undoubtedly look fainter at $z=0$ (see also \citealt{1999Helmi}; J08; \citealt{ Gomez2010}).  We explore this further in the next section.

\subsection{Mass of progenitor satellites}
\label{sec:mass}

In this section we  explore the BSPs mass distribution. In particular, we focus on the  mass of each BSP at infall, i.e. at the time of $R_{\rm vir}$ crossing (see Sec. \ref{sec:time}). We search for correlations with other properties such as their $t_{\rm infall}$ and  first pericentric distance. As we just showed, the $\mu_{\rm r}^{\rm lim}$ at which the brightest stellar streams are detected provides information about the BSP infall time. In Figure \ref{fig:dist_mass} we show the infall mass distribution for the BSPs which streams  were first detected at values of  $\mu_{\rm r}^{\rm lim} < 28.5$ (top panel) mag arcsec$^{-2}$ and $\mu_{\rm r}^{\rm lim} > 28.5$ mag arcsec$^{-2}$ (botom panel). The figure shows that BSPs with brighter LSB features are typically more massive than their fainter counterparts. The difference is of approximately 1 dex in stellar mass. 

As previously discussed, one would naively expect that the BSPs are among the most massive satellites accreted by each individual host. However satellites more massive than $10^{8} M_{\odot}$ in stellar mass are severely affected by dynamical friction and very rapidly disrupted, as discussed by F20. The more massive the satellite, the more efficient this process is. Such massive satellites tend to sink rapidly to the host galactic center, typically leaving behind shell-like low surface brightness substructures. As discussed by \citet{2015MNRAS.454.2472H} \citep[see also][]{2015MNRAS.450..575Amorisco, 2018Pop-Pipelich, 10.1093/mnras/stz1251}, this type of low surface brightness substructures tend to have shorter lifetimes than other stream types, such as loops, associated with satellites less massive than $10^{8} M_{\odot}$ in stellar mass on less eccentric orbits. Thus, BSPs may not necessarily be associated with the most massive accretion event.

In Figure~\ref{fig:time_mass} we show the distribution of BSPs infall mass against their corresponding $t_{\rm infall}$. The left and middle panels show the total and stellar mass, respectively. The symbols in both panels have been color coded according to the  $\mu_{\rm r}^{\rm lim}$ at which the brightest stream was detected. The first thing to notice is that the population of BSPs  shows a wide range of satellite masses. Values span  $9.5 \lesssim 
\log_{10} (M_{\rm tot}/M_{\odot}) \lesssim 11.5$ in total mass and $7.5 \lesssim \log_{10} (M_{*}/M_{\odot}) \lesssim 10.5$ in stellar mass. This represents a variation of two and three dex in total and stellar mass, respectively. 
In general we find brighter streams to be associated with more massive progenitors. However, a clear relation between the mass of the progenitor and its infall time can also be seen. In addition to producing the brighter streams, more massive BSPs typically have more recent infall time. Similar results were found by  J08, using dark matter-only cosmologically motivated simulations of the formation of MW stellar halos. However, unlike J08 who studied all satellites that have contributed streams at any  $\mu_{\rm r}^{\rm lim}$, we are only focusing here on the main properties of the BSPs. Therefore, our results could be potentially compared to those from the high mass end ($> 10^{7.5}$ M$_{\odot}$) of the satellite mass distribution analysed in J08, which contains satellites with stellar masses even smaller than $10^5$ M$_{\odot}$ (see their Fig. 4). Nevertheless, a direct comparison is not possible  because we cannot distinguish the brightest stream of each model from all the streams shown in J08, neither the properties of its corresponding progenitor.

As before, we subdivide the sample of BSPs by the SB at which the stream was first identified $\mu_{\rm r}^{\rm lim} = 28.5$ mag arcsec$^{-2}$. For $\mu_{\rm r}^{\rm lim} > 28.5$ mag arcsec$^{-2}$ we find BSPs with median values of  $(M_{\rm tot}, M_{*}) \approx (10^{10.4},  10^{8.8})~M_{\odot}$. On the other hand, for  $\mu_{\rm r}^{\rm lim} \leq 28.5$ mag arcsec$^{-2}$ we find  $(M_{\rm tot}, M_{*}) \approx (10^{10.9},10^{9.6})~M_{\odot}$ for the corresponding BSPs. 
The right panel of Fig.~\ref{fig:time_mass} shows the total mass -- $t_{\rm infall}$ relation, but now symbols are color coded according to whether the satellites have survived to the present-day (yellow dots) or not (blue dots). This panel clearly shows that surviving BSPs are both i) less massive at any given $t_{\rm infall}$ and ii) have been accreted later at any given $M_{\rm tot}$. The size of the symbols in this panel indicates the first satellite pericentric distance. Note that late accreted satellites that, at a given mass,  have been fully disrupted  show very small first infall pericentric distances.  Indeed, the BSPs in Au22 and Au7 have very small first pericentric distances with values of $\approx 12$ and 22 kpc, respectively. As a result, they were rapidly disrupted by strong tidal forces associated with these inner galactic regions.

\begin{figure*}
    \centering
    \includegraphics[width=\textwidth]{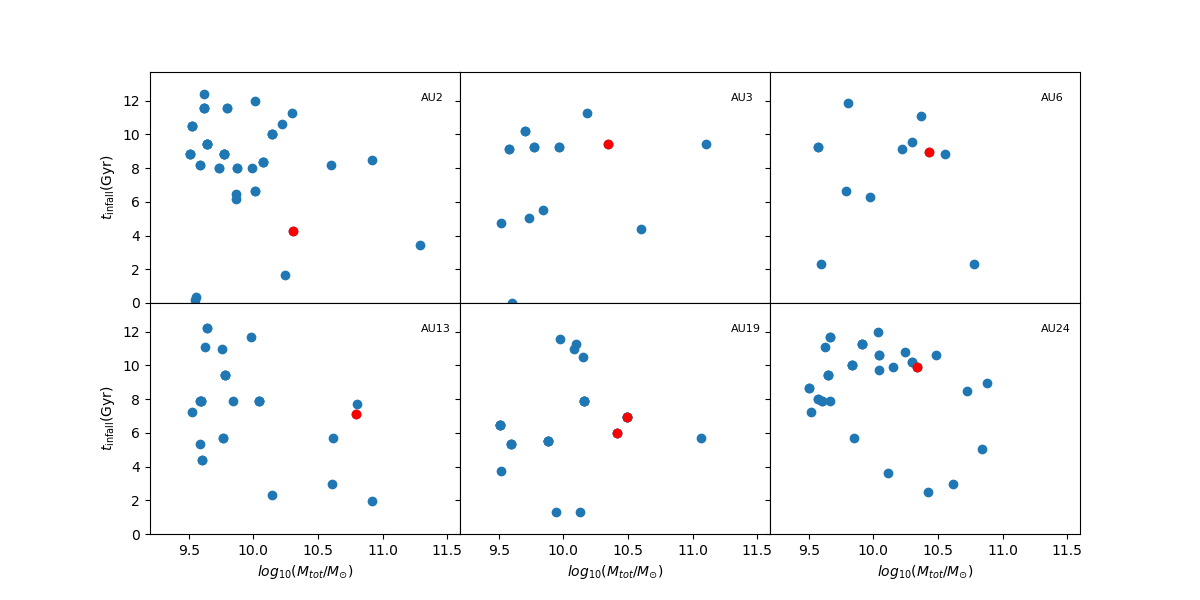}
    \caption{Each panel shows the distribution of lookback infall times and total infall mass of satellites in different Auriga halos, similar to the right panel of Fig. \ref{fig:time_mass}. Only satellites with total infall mass $> 10^{9.5}$ M$_{\odot}$ are considered. The brightest stream progenitors (BSPs) on these halos are highlighted with red dots. Note that, in all cases, the BSPs are neither the most massive nor the most recent accretion event.}
    \label{fig:mass_diff_Au}
\end{figure*}

It is interesting to highlight that in most cases BSPs are not the dominant contributor to the overall galactic stellar halos, even though they are among the significant progenitors. As shown by \citet{2019MNRAS.485.2589M}, the number of satellites that contribute  $90\%$ of the accreted halo mass (significant progenitors)  ranges from 1 to 14, with a median of 6.5 \citep[see also][F20]{cooper2010}. In Table \ref{tab:Satellite properties} we list the rank of each BSP according to its mass contributed to the overall stellar halo. Only $35\%$  of the BSPs correspond to the most significant progenitor of an Auriga halo. Another $32\%$ of the BSPs are distributed among the 2nd and 3rd most significant contributors. The remaining  $32\%$ have been ranked above the 4th significant contributor. Thus, most of these halos have accreted more massive satellites than the BSPs over time. 

This is closely related to the mass-$t_{\rm infall}$ relation shown in Fig.~\ref{fig:time_mass}. As previously discussed, BSPs accreted earlier are typically less massive than those recently accreted. The difference can be as large as two order of magnitudes in total mass. This is not purely due to the fact that at later times galaxies are more likely to accrete more massive substructures. To illustrate this, in the right panel of Fig.~\ref{fig:time_mass} we show the five most massive satellites accreted by two Auriga halos, Au5 and Au28, with red and green triangles, respectively. It is clear that both halos have accreted satellites more than 1 dex more massive than their corresponding BSPs, and at similar times.  However, as recently discussed in \citet{2021ApJ...920...10P}, the infall mass of the accreted satellites  plays a key role in  determining the mixing time scales of their streams \citep[see also][]{1998ApJ...495..297J,1999Helmi}. More massive progenitors generate warmer stellar streams due to their higher internal velocity dispersions \citep{2015MNRAS.450..575A}.  Indeed, as discussed in \citet[][]{2003MNRAS.339..834H}, streams originating in smaller haloes are narrower, more clearly defined, and, typically, they phase-mix over longer time-scales. Note however that this is not the only process at play. As shown by F20, for satellites with $M_* > 10^8 ~M_{\odot}$ (i.e. within the BSPs range), their survival time strongly depends on their  $t_{\rm infall}$. Due to dynamical friction, these more massive and luminous halos are more rapidly disrupted compared to lower mass halos accreted at similar times. In such cases, the brightest streams  can be associated with less massive (but still luminous) galaxies that can continue orbiting their host, and releasing debris, for longer periods.
As $t_{\rm infall}$  gets closer to $z=0$ the chances of finding coherent stellar streams arising from the more massive satellites grows. This is more clearly illustrated in Fig. \ref{fig:mass_diff_Au}, where we show the total mass distribution of accreted satellites (not only BSPs) as a function $t_{\rm infall}$ for six particular Auriga models. Based on the infall mass distribution of the BSPs (Fig.  \ref{fig:dist_mass} ), we only focus on accreted satellites with total infall masses $> 10^{9.5}$ M$_{\odot}$. The BSPs on each model are highlighted in red. Note that Au3 and Au19 (middle column)  show two BSPs. The reason is these halos  show, at the same $\mu_{\rm r}^{\rm lim}$, two LSB features associated with different progenitors (see Sec. \ref{sec:identify}). In all these examples the host has accreted satellites more massive than the BSPs, with later $t_{\rm infall}$. Indeed, we find that  52 percent of the models had a satellite more massive than the corresponding BSPs infalling at a later time.
As previously discussed, we find that obvious debris features observed around galaxies today are not typically associated with most recent and most luminous accretion events.

\section{Summary and Conclusions}
\label{sec:conclusion}

In this work, we have searched for the brightest stellar streams in 30 fully cosmological magneto-hydrodynamical simulations of Milky-Way mass galaxies from the Auriga project. Our main goal was to quantify the number of halos  with clear tidal streams as a function of limiting surface brightness, and to characterize the main properties of the brightest stellar stream progenitors (BSPs). The properties of the satellites within this mass range are reasonably well numerically converged at the considered resolution level. This allowed us to link, for the first time using fully cosmological hydrodynamical simulations, a clear observable property -- the brightest stream surface brightness -- with the accretion history of Milky Way-like galaxies.

For each  halo, we  generated several surface brightness (SB) maps reaching different limiting surface brightness levels. Starting from the shallower SB map, we have searched for the first clear signature of a stellar stream. This was done iteratively, by increasing the limiting surface brightness of the maps at each step. To minimize the effects that dust and the Galactic cirrus would have on our detections, we have focused our analysis on the model r-band photometry. Considering other photometric bands does not affect our main results. To take into account the effect of different galaxy inclinations on the identification of the brightest streams as a function of $\mu_{\rm r}^{\rm lim}$, we have considered edge-on and face-on disc projections. To mimic the smoothing performed in observations to enhance diffuse structure and preserve image resolution, the distribution of stellar particles was assigned to bins of $2 \times 2$ kpc, and fluxes were integrated within each bin.

None of our  models show signatures of streams for $\mu_{\rm r}^{\rm lim} \leq 25.5$ mag arcsec$^{-2}$. At the typical SB limiting magnitude reached by current surveys, $\mu_{\rm r}^{\rm lim} \approx 28.5$ mag arcsec$^{-2}$, we show that the brightest stream can be detected in $\approx$ 30 percent of our models. Independently of the projection, we find that the cumulative function of detected brightest stream strongly rises at values of 28.5 mag arcsec$^{-2}$, and then flattens again beyond 30.5 mag arcsec$^{-2}$. Furthermore, 13\% of our models show no detectable streams up to  $\mu_{\rm r}^{\rm lim}= 31$ mag arcsec$^{-2}$. Varying the projected orientation of our models has a significant impact in the detectability of bright stellar streams. With respect to the face-on projection, the cumulative function shows an increase of about 10\% of brightest stream detections at all limiting SB levels when the galaxies are projected edge-on. Our results show that, to obtain a representative census of the brightest LSB features in the nearby Universe, it is thus critical to conduct surveys that reach fainter SB limiting magnitude that current surveys ($\mu_{\rm r}^{\rm lim} \approx 28$ mag arcsec$^{-2}$, see \citealt{2013ApJ...765...28A, 2018ApJ...866..103K, 2018A&A...614A.143M}).  Such surveys will be provided by the Vera Rubin-LSST as a result of its 10 years campaign, which will reach $\mu_{\rm r}^{\rm lim} \approx 30$ mag arcsec$^{-2}$. To achieve this goal it will be key  to develop accurate methods to systematically and efficiently perform sky background subtraction, as well as to characterize sources of contamination such as Galactic Cirrus \citep[see e.g.,][]{2020A&A...644A..42R}, which become relevant for observation deeper than $\mu_{\rm r}^{\rm lim} \approx 28.5$ mag arcsec$^{-2}$.

In general we find our results are in good agreement with previous observational studies. We note however that the comparison between our results and those presented in, e.g., \citet{2013ApJ...765...28A} and \citet{2018A&A...614A.143M} is not straightforward, and there are several  differences to take into account. First, our models do not account for dust extinction and background noise, both likely to conceal signatures of faint stellar substructures. We also  have a much smaller sample of galaxies which, on average, are more luminous than the observed ones. 
Because of these differences, it is expected the fraction of galaxies with streams detected in the models to be slightly larger than those found in observations, which are likely to represent a lower detection limit. Finally it is worth noting that, even though the properties of the simulated satellite populations are generally in good agreement with those observed in the MW and M31 satellites, our results are still limited by certain aspects of the model, such as numerical resolution and stellar feedback. For example, while the satellite phase-space scaling relation is well reproduced, the detailed internal structure of our simulated satellites, specially those with lower mass, are likely to be unresolved. In addition, diffusion of stellar streams in phase-space could also be enhanced due to the limited numerical resolution. Note that since we are focusing on the brightest stellar streams on each model, this limitation should be minimized. We also note that, as reported in \citet{2019MNRAS.485.2589M}, Auriga stellar halos are slightly more massive than some of the observed halos in the Local Universe. This is partially the result of the accretion of slightly more massive satellites than observed \citep{2018MNRAS.478..548S}.

We have identified and characterized the main properties of the brightest stream progenitors (BSPs), focusing in particular on their infall times (the time at which they first cross the virial radius of the host galaxy) and their infall mass. We find that BSPs can be accreted in a very wide range of $t_{\rm infall}$, with values that can range from 10 Gyr ago to very recent accretion events at $t_{\rm infall} = 1.6$ Gyr. Interestingly, only $25$ percent of the BSPs have been recently accreted, within the last 5 Gyr. Thus, most BSPs correspond to relatively early accretion events (i.e., $5 < t_{\rm infall} \lesssim 10$ Gyr). As expected, BSPs associated with brighter stellar streams ($\mu_{\rm r}^{\rm lim} < $ mag arcsec$^{-2}$) were typically accreted later than those with fainter substructures ($\mu_{\rm r}^{\rm lim} > 28.5$ mag arcsec$^{-2}$). We also find that only 37 percent of the BSPs can still be identified at the present-day. The median $t_{\rm infall}$ for the  surviving and the disrupted BSPs are $t_{\rm infall} = 5.2$ and 6.7 Gyr, respectively. 

Looking at the BSPs infall mass, we find a wide range of masses with values with $9.5 \lesssim$ $\log_{10} M_{\rm tot}/M_\odot \lesssim 11.5$  in total mass and $7.5 \lesssim$ $\log_{10}M_{*}/M_\odot \lesssim 10.5$ in stellar mass. This represents a variation of 2 and 3 dex in total and stellar mass, respectively. A comparison between surviving and destroyed BSPs shows that surviving BSPs are less massive at any given $t_{\rm infall}$ and have been accreted later at any given $M_{\rm tot}$. We find that brighter streams tend be associated with more massive BSPs. Indeed, there is a correlation between the BSPs infall mass and $t_{\rm infall}$, such that more massive progenitors tend to be accreted at later times. However, we showed that this is not simply due to the fact that close to  $z=0$ galaxies are more likely to accrete more massive substructures.  Indeed, we find that haloes that have relatively low mass BSPs accreted at earlier times have accreted other satellites up to 1 dex more massive than the BSPs, and at similar times. Due to their larger internal velocity dispersion, debris from these more massive and luminous satellites are more rapidly phase-mixed. Furthermore, dynamical friction acts more efficiently on these more massive and luminous halos. As a result, they are more rapidly disrupted compared to lower mass halos accreted at similar times. In such cases, the brightest streams can be associated with less massive (but still luminous) galaxies that can continue orbiting their host, and realising debris, for longer periods. Finally, we also show that, for most of the cases, BSPs are not the dominant contributors to the accreted stellar halo, even though they are always significant contributors; i.e. part of the subset of satellites that contribute over 90 percent of the accreted stellar halo mass. 
Our work thus complements the information that can be obtained from stellar halos regarding the accretion history of late-type MW-mass galaxies \citep[e.g.,][]{2016ApJ...821....5D,2018NatAs...2..737D, 2019MNRAS.485.2589M}, namely the dominant accretion event, by adding extra information provided by the stellar streams.

\section*{Acknowledgements}

We wish to thank the anonymous reviewers for thoughtful and constructive reports that helped us improve our manuscript. AV and FAG acknowledges support from ANID FONDECYT Regular 1211370. AV acknowledges support from DIDULS PTE192137. AM acknowledges support from ANID FONDECYT Regular 1212046. FAG and AM gratefully acknowledge support by the ANID BASAL project FB210003. FAG, AM, DP and IG acknowledge funding from the Max Planck Society through a “Partner Group” grant. IG acknowledges financial support from CONICYT Programa de Astronom\'ia, Fondo ALMA-CONICYT 2017 31170048. FM acknowledges support through the Program "Rita Levi Montalcini" of the Italian MUR. CSF acknowledges support from European Research Council (ERC) Advanced Investigator grant DMIDAS (GA 786910). This work was also supported by the Consolidated Grant for Astronomy at Durham (ST/L00075X/1). This work made use the DiRAC Data Centric system at Durham University, operated by the Institute for Computational Cosmology on behalf of the STFC DiRAC HPC Facility (www.dirac.ac.uk). This equipment was funded by BIS National E-infrastructure capital grant ST/K00042X/1, STFC capital grants ST/H008519/1 and ST/K00087X/1, STFC DiRAC Operations grant ST/K003267/1 and Durham University. DiRAC is part of the National E-Infrastructure

\section*{DATA AVAILABILITY}

The data underlying this article will be shared on reasonable request to the corresponding author.



\bibliographystyle{mnras}
\bibliography{main_text} 






\bsp	
\label{lastpage}
\end{document}